\RequirePackage{fix-cm}

\documentclass{article}

\usepackage{stmaryrd}%\llbracket etc.
\usepackage{rotating}
\usepackage{url}
\usepackage{bm}
\usepackage[basic]{complexity}
\usepackage{ifthen}%managing multiple versions
\newboolean{short_version}% for proceedings
\newboolean{arxiv_version}% full version with proofs
\usepackage[pdflatex,recompilepics=true]{gastex}
\usepackage{authblk}
\usepackage{mathtools}%for \shortintertext
\usepackage{fullpage}
\newcounter{theorem}
\usepackage{amsmath,amsfonts,amssymb,amsthm}
\newtheorem{thm}[theorem]{Theorem}{\bf}{\it}
{\bf}{\it}
\newtheorem{defn}[theorem]{Definition}{\bf}{\it}
{\bf}{\it}
\newtheorem{cor}[theorem]{Corollary}{\bf}{\it}
\newtheorem{lem}[theorem]{Lemma}{\bf}{\it}
{\it}{\rm}
\newtheorem{rem}[theorem]{Remark}{\it}{\rm}
\newcommand{\PEPd}{\PEP_{\mathrm{dir}}}
\newcommand{\PEPdcd}{\PEP_{\mathrm{co\&dir}}}
\newcommand{\PEPpd}{\PEP^{\mathrm{partial}}_{\mathrm{dir}}}
\newcommand{\PEPpcod}{\PEP^{\mathrm{partial}}_{\mathrm{codir}}}

\newcommand{\PEPdcflpcod}{\PEP^{\mathrm{partial[DCFL]}}_{\mathrm{codir}}}

\newcommand{\PEPprespcod}{\PEP^{\mathrm{partial[Pres]}}_{\mathrm{codir}}}

\newcommand{\bfF}{{\textbf{F}}}%bold F, for ordinal-recursive complexity
\newcommand{\tta}{{\mathtt{a}}}%a in tt for letters
\newcommand{\ttb}{{\mathtt{b}}}%b in tt for letters
\newcommand{\ttc}{{\mathtt{c}}}%c in tt for letters
\newcommand{\ttz}{{\mathtt{z}}}%c in tt for letters
\newcommand{\inl}{{\,\mid\mid\mid\,}}

\newcommand{\xs}{{\mathtt{\dag}}}%xs = eXtra Symbol. I'm not yet fixed on what to use
\newcommand{\btr}{{\blacktriangleright}}
\newcommand{\btl}{{\blacktriangleleft}}

\newcommand{\ubracew}[2]{{\underset{#2}{\underbrace{#1}}}}
\newcommand{\obracew}[2]{{\overset{#2}{\overbrace{#1}}}}

\newcommand{\Nat}{{\mathbb{N}}} % use the right fonts
\renewcommand{\emptyset}{\varnothing}% looks better
\renewcommand{\epsilon}{\varepsilon} % we want all epsilons the same
\newcommand{\eps}{\varepsilon} % we want all epsilons the same
\renewcommand{\setminus}{\smallsetminus} % we want all setminus the same
\newcommand{\overto}[1]{\xrightarrow{\!\!#1\!\!}}
\newcommand{\step}[1]{\overto{#1}} % handy synonym
\newcommand{\rew}{{\step{}_{\Delta}}}
\newcommand{\rewn}[1]{{\step{#1}_{\Delta}}}
\newcommand{\rews}{{\step{*}_{\Delta}}}

\newcommand{\egdef}{\stackrel{\text{{\tiny def}}}{=}}
\newcommand{\equivdef}{\stackrel{\text{{\tiny def}}}{\Leftrightarrow}}

\newcommand{\mirror}[1]{\widetilde{#1}}

\newcommand{\size}[1]{{\mathopen{\mid}#1\mathclose{\mid}}}

\newcommand{\subword}{\sqsubseteq}

\renewcommand{\qed}{}

\input{just-the-cites.aux}%ugly hack so that \ref's in gpicture's work with [pdflatex,recompilepics=true]{gastex}

\begin{document}

\title{Generalized Post Embedding Problems\thanks{
Supported  by 
Grant ANR-11-BS02-001. The first author was partially supported by Tata Consultancy Services. An extended abstract of this article
    appeared in~\cite{KS-csr2012}.}  
}

%\author{P. Karandikar \and Ph. Schnoebelen}

%\institute{P. Karandikar \at Chennai Mathematical Institute and LSV, ENS Cachan
%           Ph.\ Schnoebelen \at
%        LSV -- CNRS \& ENS Cachan %\\ \url{www.lsv.ens-cachan.fr/~phs}
%}

\author[1,2]{P. Karandikar}
\author[1]{Ph. Schnoebelen}
\affil[1]{LSV -- CNRS \& ENS Cachan}
\affil[2]{Chennai Mathematical Institute}

\date{}

\maketitle

%%% File: abstract.tex -*-LaTeX-*-

\begin{abstract}
The Regular Post Embedding Problem extended with partial (co)di\-rect\-ness is
shown decidable. This extends to universal and/or counting versions. 
It is also shown that combining directness and codirectness in Post
Embedding problems leads to undecidability.
\end{abstract}

%%% Local Variables:
%%% fill-column: 75
%%% ispell-check-comments: nil
%%% Local IspellDict: "english"
%%% End:

%%% Local Variables:
%%% fill-column: 75
%%% ispell-check-comments: nil
%%% Local IspellDict: "english"
%%% End:

\setcounter{tocdepth}{3}
%\tableofcontents % while preparing

%%% File: sec-intro.tex -*-LaTeX-*-

\section{Introduction}
%=====================
\label{sec-intro}

The \emph{Regular Post Embedding Problem} ($\PEP$ for short, named by
analogy with Post's Correspondence Problem, aka $\PCP$) is the problem of deciding,
given two morphisms on words $u,v:\Sigma^*\to \Gamma^*$ and a regular
language $R\in\Reg(\Sigma)$, whether there is $\sigma \in R$ such that
$u(\sigma)$ is a (scattered) subword of $v(\sigma)$. One then calls
$\sigma$ a \emph{solution} of the $\PEP$ instance.

We use ``$\subword$''
to denote the \emph{subword}
relation, also called \emph{embedding}:
$u(\sigma)\subword v(\sigma)$ $\equivdef$ $u(\sigma)$ can be obtained by
erasing some letters from $v(\sigma)$, possibly all of them, possibly none.
Equivalently, $\PEP$ is the question whether a rational relation, or a
transduction, $T\subseteq\Gamma^*\times\Gamma^*$ intersects non-vacuously
the subword relation~\cite{barcelo2013}, hence it is a special case of the intersection problem
for two rational relations.

This problem, introduced in~\cite{CS-fsttcs07}, is new and quite remarkable: it is
decidable but surprisingly hard since it is not
primitive-recursive.\footnote{But the
  problem becomes easy, decidable in linear-time and logarithmic
  space~\cite{CS-fsttcs07}, when restricted to $R=\Sigma^+$ as in $\PCP$.}
The problem is in fact $\bfF_{\omega^\omega}$-complete~\cite{KS-fossacs2013}, that is, it sits at
the first level above multiply-recursive in the
Ordinal-Recursive Complexity Hierarchy~\cite{schmitz2013}.

A variant problem was introduced in~\cite{CS-fsttcs07}: $\PEPd$ asks for
the existence of a \emph{direct}
solution, i.e., some $\sigma\in R$ such that $u(\tau)\subword
v(\tau)$ for every prefix $\tau$ of $\sigma$. It turns out that $\PEP$ and
$\PEPd$ are
inter-reducible (though not trivially)~\cite{CS-omegapep} and have the same complexity.
\\

In this article
we introduce $\PEPpd$, or ``$\PEP$ with \emph{partial} directness'':
Instead of requiring $u(\tau)\subword v(\tau)$ for all prefixes of a
solution (as in $\PEPd$), or for none (as in $\PEP$), $\PEPpd$ lets us
select, by means of a regular language, which prefixes should verify the
requirement. Thus $\PEPpd$ generalizes both $\PEP$ and $\PEPd$.

Our main result is that $\PEPpd$ and the mirror problem $\PEPpcod$ are decidable.
The proof combines two ideas. Firstly, by
Higman's Lemma, a long solution must eventually contain
``\emph{comparable}'' so-called cutting points, from which one deduces that
the solution is not minimal (or unique, or \ldots). Secondly, the above
notion of ``\emph{eventually}'', that comes from Higman's Lemma, can be
turned into an effective upper bound thanks to a Length Function Theorem~\cite{SS-icalp11}.

The decidability of $\PEPpd$ not only generalizes the decidability of
$\PEP$ and $\PEPd$: it is also simpler than the earlier proofs for $\PEP$
or $\PEPd$, and it easily leads to an $\bfF_{\omega^\omega}$ complexity
upper bound.

In a second part of the article, we extend our main result and show the
decidability of universal and/or counting versions of the extended $\PEPpd$
problem. We also explain how our attempts at further generalisation, most
notably by considering the combination of directness and codirectness in a
same instance, lead to undecidable problems.

\paragraph{Applications to channel machines.}
%---------------------------------------
Our interest in $\PEP$ and
its variants comes from their close connection with fifo channel
machines, a family of computational models that play  a
central role in some areas of program and system verification
(see~\cite{cece95,abdulla-forward-lcs,latorre2008,atig2008}) and that also provide
decidable automata models for
 problems on Real-Time and Metric Temporal Logic, modal
logics, data logics,
etc.~\cite{abdulla-icalp05,ouaknine2006,kurucz06,konev06,lasota2008,barcelo2013}.
Here, $\PEP$ and its variants
provide abstract versions of problems on channel
machines, bringing greater clarity and versatility in
both decidability and undecidability (more generally, hardness) proofs.

\begin{figure}[htbp]
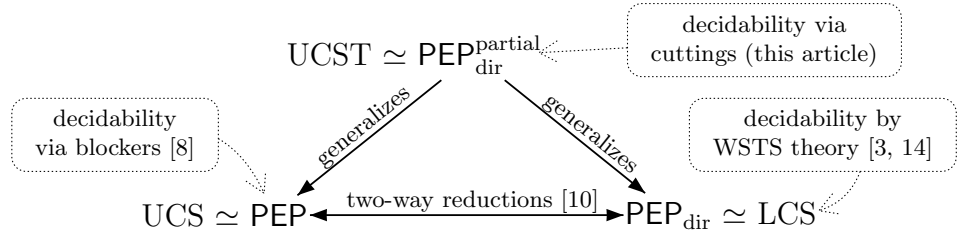

\centering
{\setlength{\unitlength}{0.97mm} % relevant for width computation with \ubracew{}
\begin{gpicture}(122,28)(0,5)
%\put(0,5){\framebox(122,28){}}

{
% TOP NODES
\node[Nw=16,Nh=7,Nframe=n](ucst)(41.33,28){}%\large UCST}
\node[Nw=18,Nh=7,Nframe=n](peppd)(60.07,28){}%\large $\PEPpd$}
\node[Nw=99,Nh=7,Nframe=n](ucst=peppd)(52,28){\large UCST $\simeq$ $\PEPpd$}

% BOTTOM LEFT NODES
\node[Nw=10.2,Nh=6,Nframe=n](ucs)(19.4,6){}%\large UCS}
\node[Nw=10.2,Nh=6,Nframe=n](pep)(32.6,6){}%\large $\PEP$}
\node[Nw=99,Nh=6,Nframe=n](ucs=pep)(26,6){\large UCS $\simeq$ $\PEP$}

% BOTTOM RIGHT NODES
\node[Nw=13.4,Nh=6,Nframe=n](pepd)(87.4,6){}%\large $\PEPd$}
\node[Nw=10.2,Nh=6,Nframe=n](lcs)(102.4,6){}%\large LCS}
\node[Nw=99,Nh=6,Nframe=n](pepd=lcs)(94,6){\large $\PEPd$ $\simeq$ LCS}
}

{\gasset{dash={0.2 0.35}0,AHnb=1,AHLength=3,AHlength=0,AHangle=30,Nmr=2}
 \node[Nadjust=wh](pf3)(100,30){\small \begin{tabular}{c}decidability via\\cuttings (this article)\end{tabular}}
 \node[Nadjust=wh](pf1)(108,17){\small \begin{tabular}{c}decidability by\\WSTS theory~\cite{abdulla96b,finkel98b}\end{tabular}}
 \node[Nadjust=wh](pf2)(11,17){\small \begin{tabular}{c}decidability\\via blockers~\cite{CS-fsttcs07}\end{tabular}}
\drawedge[curvedepth=0,exo=4](pf3,peppd){}
\drawedge[curvedepth=6](pf2,pep){}
\drawedge[curvedepth=4,sxo=6](pf1,lcs){}
}

{\gasset{linewidth=0.25,AHLength=2.5,AHlength=2.3,ATLength=2.5,ATlength=2.3} %lines for reductions
\drawedge[ATnb=1,ELside=r,ELdist=0.3](pepd,pep){\small two-way reductions~\cite{CS-omegapep}}
\drawedge[ELside=r,ELdist=0.9,ELpos=73](peppd,pep){\begin{rotate}{39}{\small generalizes}\end{rotate}}
\drawedge[ELside=l,ELdist=0.9,ELpos=33](peppd,pepd){\begin{rotate}{-39}{\small generalizes}\end{rotate}}
}
\end{gpicture}
}
\caption{Three decidability proofs for $\PEP$ and variants}
\label{fig-roadmap}
\end{figure}

In this context, a further motivation for considering $\PEPpd$ is that it allows
solving the decidability of UCSTs, i.e., unidirectional channel systems
(with one reliable and one lossy channel) \emph{extended with the
  possibility of testing the contents of channels}~\cite{JKS-tcs2012}. We
recall that $\PEP$ was introduced for UCSs, unidirectional channel systems
where tests on channels are not supported~\cite{CS-omegapep,CS-concur08},
and that $\PEPd$ corresponds to LCSs, i.e., lossy channel systems, for which
 decidability use techniques from WSTS
theory~\cite{abdulla96b,finkel98b,CS-lics08,BS-fmsd2013}.
Fig.~\ref{fig-roadmap} depicts the resulting situation.
\\
%
%%% File: fig-roadmap.tex -*-LaTeX-*-
%%%
%
%
%%% Local Variables:
%%% fill-column: 130
%%% ispell-check-comments: nil
%%% Local IspellDict: "english"
%%% End:

%

\paragraph{Outline of the article.}
%================================
Section~\ref{sec-defn} recalls basic notations and definitions. In
particular, it lists basic results about how the subword relation
interacts with concatenations and factorization.
Section~\ref{sec-higman} explains the Length Function Theorem for Higman's
Lemma.
Section~\ref{sec-peprpd} contains our main result, a direct decidability
proof for $\PEPpd$, a problem subsuming both $\PEP$ and $\PEPd$.
Section~\ref{sec-infpeprpd} builds on this result and shows the
decidability of counting problems on $\PEPpd$.
Section~\ref{sec-universal} further shows the decidability of universal
variants of these questions.
Section~\ref{sec-undec} contains undecidability results for
some extensions of $\PEPpd$.
%

%%% Local Variables:
%%% fill-column: 75
%%% ispell-check-comments: nil
%%% Local IspellDict: "english"
%%% End:

% LocalWords:  UCS LCS codirectness XPath UCZ UCST UCSTs UCSs LCSs

%%% File: sec-defn.tex -*-LaTeX-*-

\section{Words and subwords}
%===========================
\label{sec-defn}

\paragraph{Words.}
%=================
Concatenation of words is denoted multiplicatively, with $\epsilon$
denoting the empty word. We write $\size{s}$ for the length of a word $s$,
and $\size{\Gamma}$ for the size of a finite alphabet $\Gamma$.
If $s$ is a prefix of a word $t$, $s^{-1}t$ denotes the unique word $s'$ such that
$t = ss'$ (otherwise $s^{-1}t$ is not defined).
Similarly, when $s$ is a suffix of $t$, $ts^{-1}$ is $t$ with the $s$
suffix removed.
%Positions in a word of length $n$ are indexed from $0$ to $n-1$.
For a
word $s = \tta_0\ldots \tta_{n-1}$, $\mirror{s} \egdef \tta_{n-1}\ldots
\tta_0$ is the mirrored word. The mirror of a language $R$ is $\mirror{R} \egdef \{ \mirror{s} ~|~ s \in R \}$.

With a language $R\subseteq\Gamma^*$ one associates a congruence (wrt
concatenation) given by $s \sim_R t \equivdef \forall x,y ( x s y \in R
\Leftrightarrow x t y \in R )$ and called the Myhill congruence (also, the
syntactic congruence). This equivalence has finite index if (and only if)
$R$ is regular. For regular $R$, let $\mu(R)$ denote this index: it
satisfies $\mu\bigl(\mirror{R}\bigr)=\mu(\Gamma^*\setminus R)=\mu(R)$ and
$\mu(R\cap R')\leq \mu(R)\,\mu(R')$. Also, $\mu(R)$ is computable from $R$,
and in particular, $\mu(R)\leq m^m$ when $R$ is recognized by a $m$-state
complete DFA~\cite{holzer2004}.

\paragraph{Subwords.}
%====================
We write $s\subword t$ when $s$ is a subword (subsequence) of $t$.
Formally, $\tta_0 \ldots \tta_{n-1}\subword t$ iff $t$ is some
concatenation $t_0 \tta_0 t_1 \tta_1 \ldots \tta_{n-1} t_n$. 
An \emph{embedding} of $s=\tta_0 \ldots \tta_{n-1}$ into $s'= \tta'_0
\ldots\tta'_{m-1}$ is a strictly monotonic map
$h:\{0,\ldots,n-1\}\to\{0,\ldots,m-1\}$ such that $\tta_i=\tta'_{h(i)}$ for
all $0\leq i<n$. Clearly, $s\subword s'$ iff there exists an embedding of
$s$ into $s'$.

The subword
relation is an ordering that is compatible with the monoid structure:
$\epsilon\subword s$ for all words $s$, and $s s'\subword t t'$ when
$s\subword t$ and $s'\subword t'$.

\begin{lem}[Subwords and concatenation]
\label{basic}
For all words $y,z,s,t$:
\begin{description}
\item[(a)]
If $y z \subword s t$, then $y \subword s$ or $z \subword t$.
\item[(b)]
If $y z \subword s t$ and $z \subword t$ and $x$ is the longest suffix of
$y$ such that $x z \subword t$, then $y x^{-1} \subword s$.
\item[(c)]
If $y z \subword st$ and $z \not \subword t$ and $x$ is the shortest prefix
of $z$ such that $x^{-1} z \subword t$, then $y x \subword s$.
\item[(d)]
If $y z \subword s t$ and $z \subword t$ and $x$ is the longest prefix of
$t$ such that $z \subword x^{-1} t$, then $y \subword s x$.
\item[(e)]
If $y z \subword s t$ and $z \not \subword t$ and $x$ is the shortest
suffix of $s$ such that $z \subword x t$, then $y \subword s x^{-1}$.
\item[(f)]
If $s x \subword y t$ and $t \subword s$, then $s x^k \subword y^k t$
for all $k \geq 1$.
\item[(g)]
If $x s \subword t y$ and $t \subword s$, then $x^k s \subword t y^k$ for
all $k \geq 1$.
\end{description}
\end{lem}
\begin{proof}
Items {(a--e)} are easy (or see~\cite[Section~3]{CS-countpep}).
Item {(f)} is proved by induction on $k$. The claim is true for $k=1$, suppose
it is true for $k=p$. Then $s x^{p+1} = s x^p x \subword y^p t x \subword
y^p s x \subword y^p y t = y^{p+1}t$.
Item {(g)} is obtained from {(f)} by mirroring.
\qed
\end{proof}

\section{Higman's Lemma and the length of bad sequences}
%=======================================================
\label{sec-higman}

It is well-known that for words over a finite alphabet,
$\subword$ is a well-quasi-ordering, that is, any infinite sequence of
words $x_1, x_2, x_3, \ldots$ contains an infinite increasing subsequence
$x_{i_1}\subword x_{i_2}\subword x_{i_3}\subword \cdots$~\cite{kruskal72}.
This result is called Higman's Lemma.

For $n\in\Nat$, we say that a sequence (finite or infinite) of words is
\emph{$n$-good} if it contains an increasing subsequence of length $n$. It is \emph{$n$-bad}
otherwise. Higman's Lemma states that every infinite sequence is $n$-good
for every $n$. Hence every $n$-bad sequence is finite.

It is often said that Higman's Lemma is ``non-effective'' or
``non-cons\-truc\-tive'' since it does not come with any explicit
information on the maximal length of bad sequences. Consequently, when one
uses Higman's Lemma to prove that an algorithm terminates, no meaningful
upper-bound on the algorithm's running time is derived from the proof.
However, the length of bad sequences can be bounded if one takes into
account the complexity of the sequences, or more precisely, of the process
that generates bad sequences. The interested reader can
consult~\cite{SS-icalp11,SS-esslli2012} for more details. In this article
we only use the simplest version of these results, i.e., the statement that
when sequences only grow in a restricted way then the maximal length of bad
sequences is computable, as we now explain.
\\

For $k\in\Nat$, we say that a sequence of words $x_1, x_2, \ldots$ is
\emph{$k$-controlled} if $\size{x_i} \leq ik$ for all $i=1,2,\ldots$ Let
$H(n,k,\Gamma)$ be the maximum length (if it exists) of an $n$-bad
$k$-controlled sequence of words over a finite alphabet $\Gamma$.
\begin{thm}[Length Function Theorem]
$H$ is a computable (total) function. Furthermore, $H$ is monotonic
in its three arguments.
\end{thm}
\begin{proof}
Any prefix of a finite $k$-controlled $n$-bad sequence is $k$-controlled
and $n$-bad. In particular, the empty sequence is. We arrange the set of
all finite $k$-controlled $n$-bad sequences into a tree denoted
$T_{n,k,\Gamma}$, or simply $T$, where the empty sequence is the root of
$T$, and where a non-empty sequence of the form $x_1,\ldots,x_{l+1}$ is a
child of its immediate prefix $x_1,\ldots,x_l$.

If $T$ has an infinite path, this path is a chain of finite bad sequences
linearly ordered by the prefix ordering and with which we can build an
infinite $k$-controlled $n$-bad sequence by taking a limit. Thus $T$ has no
infinite paths since, by Higman's Lemma, $\Gamma^*$ has no infinite bad
sequences. Furthermore $T$ is finitely branching, since the sequences it
contains are $k$-controlled and $\Gamma$ is finite. Thus, by K\H{o}nig's
Lemma, $T$ is finite and $H(n,k,\Gamma)$ exists: it is the length of the
longest sequence appearing in $T$, and also the length of $T$'s longest
path from the root.

$H$ is computable since $T_{n,k,\Gamma}$ can be constructed effectively,
starting from the root and listing the finitely many ways a current $n$-bad
sequence can be extended in a $k$-controlled way. Finally, $H$ is monotonic
since, when $n'\leq n$ and $k'\leq k$, the $n$-bad $k$-controlled sequences
over $\Gamma$ include in particular all the $n'$-bad $k'$-controlled
sequences over a subalphabet.
\qed
\end{proof}
\begin{rem}
Note that there is in general no maximum length of $n$-bad sequences over
$\Gamma$ if one does not restrict to {$k$-controlled} sequences.
However, the proof of the Length Function Theorem can accommodate more
liberal notions of controlled sequences, e.g., having $\size{x_i}\leq f(i)$
for all $i$, where $f$ is a given
computable function.

Note also that if $\size{\Gamma}=\size{\Gamma'}$ then
$H(n,k,\Gamma)=H(n,k,\Gamma')$: only the number of different letters in
$\Gamma$ matters, and we sometimes write $H(n,k,p)$ for
$H(n,k, \Gamma)$ where $p=\size{\Gamma}$.
Upper bounds on $H(n,k,p)$ can be derived from the results given
in~\cite{SS-icalp11} but these bounds are enormous, hard to express and
hard to understand. In this article we content ourselves with the fact that
$H$ is computable.
\qed
\end{rem}
Below, we use the Length Function Theorem contrapositively: a
$k$-controlled sequence of length greater than $H(n, k, {\Gamma})$ is
necessarily $n$-good, i.e., contains an increasing subsequence
$x_{i_1}\subword x_{i_2}\subword \cdots \subword x_{i_n}$ of length $n$.

%%% Local Variables:
%%% fill-column: 75
%%% ispell-check-comments: nil
%%% Local IspellDict: "english"
%%% End:

% LocalWords:  nig's ss ik fsttcs contrapositively Myhill

%%% File: sec-peprpd.tex -*-LaTeX-*-

\section{Deciding $\PEPpd$, or $\PEP$ with partial directness}
%=============================================================
\label{sec-peprpd}

We introduce $\PEPpd$, a problem  generalizing both $\PEP$ and $\PEPd$,
and show its decidability. This is proved by showing that if a
$\PEPpd$ instance has a solution, then it has a solution whose length is
bounded by a computable function of the input. This is simpler and
more direct than
the earlier decidability proof (for $\PEP$ only) based on
blockers~\cite{CS-fsttcs07}.

\begin{defn}
$\PEPpd$ is the problem of deciding, given morphisms $u,v : \Sigma^*
\to \Gamma^*$ and regular languages $R, R' \in\Reg(\Sigma)$, whether
there is $\sigma \in R$ such that $u(\sigma) \subword v(\sigma)$ and
$u(\tau) \subword v(\tau)$ for all prefixes $\tau$ of $\sigma$  belonging
to $R'$ (in which case $\sigma$ is called a \emph{solution}).
\\
$\PEPpcod$ is the variant problem of deciding whether there is
$\sigma\in R$ such that $u(\sigma)\subword v(\sigma)$ and $u(\tau)\subword
v(\tau)$ for all suffixes $\tau$ of $\sigma$ that belong to $R'$.
\end{defn}
Both $\PEP$ and $\PEPd$ are special cases of $\PEPpd$, obtained by taking
$R' = \emptyset$ and $R' = \Sigma^*$ respectively. Obviously $\PEPpd$ and
$\PEPpcod$ are two equivalent presentations, modulo mirroring, of a same
problem.
Given a $\PEPpd$ or $\PEPpcod$ instance, we
let $K_u\egdef\max_{a\in\Sigma}\size{u(a)}$ denote the \emph{expansion
  factor} of $u$ and
 define 
\[
                   L\:\egdef\: H(\mu(R) \, \mu(R') + 1, K_u, \Gamma)
\:
\]
(recall that $\mu(R)$ and $\mu(R')$ are the indexes of the Myhill
congruences associated with $R$ and $R'$, while $H(n,k,\Gamma)$ is defined
with the Length Function Theorem).

In this section we prove:
\begin{thm}
\label{theo-short}
A $\PEPpcod$ instance has a solution if, and only if, it has a
solution of length at most $2L$.
\\
This entails that $\PEPpcod$ is decidable.
\end{thm}
Decidability is an obvious consequence since the length bound is
computable, and since it is easy to check whether a candidate $\sigma$ is a
solution. 

For the proof of Theorem~\ref{theo-short}, we 
consider an arbitrary
$\PEPpcod$ instance $(\Sigma, \Gamma, u,
v, R, R')$ and a solution $\sigma$. Write $N=\size{\sigma}$ for
its length, $\sigma[0,i)$ and $\sigma[i,N)$ for, respectively, its prefix
of length $i$ and its suffix of length $N-i$. Two indices
$i,j\in [0,N]$ are \emph{congruent} if $\sigma[i,N) \sim_R
\sigma[j,N)$ and $\sigma[i,N) \sim_{R'} \sigma[j,N)$. When $\sigma$
is fixed, as in the rest of this section, we use
shorthand notations like $u_{0,i}$ and $v_{i,j}$ to denote the
images, here $u(\sigma[0,i))$ and $v(\sigma[i,j))$, of factors
of $\sigma$.

We prove two ``cutting lemmas'' giving sufficient conditions for
``cutting'' a solution $\sigma = \sigma[0,N)$ along  certain indices $a <
  b$, yielding a shorter solution $\sigma' = \sigma[0,a)\sigma[b,N)$, i.e.,
      $\sigma$ with the factor $\sigma[a,b)$ cut out.
Here
the following notation is useful. We associate, with every suffix $\tau$ of
$\sigma'$, a corresponding suffix, denoted $S(\tau)$, of $\sigma$: if $\tau$ is a suffix of $\sigma[b,N)$, then $S(\tau) \egdef
\tau$, otherwise, $\tau = \sigma[i,a)\sigma[b,N)$ for some $i<a$ and we let
$S(\tau) \egdef \sigma[i,N)$. In particular $S(\sigma') = \sigma$.

%% G O O D => blue / B A D => red
An index $i \in [0,N]$ is said to be \emph{blue} if $u_{i,N} \subword
v_{i,N}$, it is \emph{red} otherwise. In particular, $N$ is blue trivially,
$0$ is blue since $\sigma$ is a solution, and $i$ is blue whenever
$\sigma[i,N) \in R'$.
If $i$ is a blue
index, let $l_i\in\Gamma^*$ be the longest suffix of $u_{0,i}$ such that
$l_i \, u_{i,N} \subword v_{i,N}$ and call it the \emph{left margin} at $i$.
\begin{lem}[Cutting lemma for blue indices]
\label{cutblue}
Let $a < b$ be two congruent and blue indices. If $l_a \subword l_b$, then
$\sigma' = \sigma[0,a)\sigma[b,N)$ is a solution (shorter than $\sigma$).
\end{lem}
\iffalse % SKIP BUT DONT REMOVE
\begin{center}
\begin{figure}
\begin{gpicture}(110,35)(0,5)
\small
\gasset{AHnb=0,Nframe=n}
\drawline(10,10)(100,10)
\drawline(10,40)(100,40)
\node(text)(6,10){$u(\sigma)$}
\node(text)(6,40){$v(\sigma)$}
\drawline(40,10)(40,8) \drawline(70,10)(70,8)
\drawline(40,40)(40,38) \drawline(70,40)(70,38)
\drawline(35,10)(35,9)\drawline(60,10)(60,9)
\node(text)(41,5){$a$}
\node(text)(71,5){$b$}
\node(text)(37.5,8){\scriptsize $l_a$}
\node(text)(65,8){\scriptsize $l_b$}
\drawline[dash={1 1}](35,10)(40,40)
\drawline[dash={1 1}](60,10)(70,40)
\end{gpicture}
\end{figure}
\end{center}
\fi%END OF SKIPPED PART
\begin{proof}
Clearly $\sigma' \in R$ since $\sigma\in R$ and $a$ and $b$ are congruent.
Also, for all suffixes $\tau$ of $\sigma'$, $S(\tau) \in R'$ iff $\tau \in R'$.

We claim that, for any suffix $\tau$ of $\sigma'$, if $u(S(\tau)) \subword
v(S(\tau))$ then $u(\tau) \subword v(\tau)$. This is obvious when
$\tau=S(\tau)$, so we assume $\tau \neq S(\tau)$, i.e., $\tau =
\sigma[i,a)\sigma[b,N)$ and $S(\tau) = \sigma[i,N)$ for some $i<a$.
Assume $u(S(\tau))\subword v(S(\tau))$, i.e.,
 $u_{i,N} \subword v_{i,N}$.
Now both $u_{i,a}$ and $l_a$ are suffixes of $u_{0,a}$, so that one is a suffix of the other, which
gives two cases. 

1.\ If $u_{i,a}$ is a suffix of $l_a$, then
\begin{xalignat*}{2}
u(\tau) = u_{i,a} \, u_{b,N} & \subword l_a \, u_{b,N} 
                &&\text{since $u_{i,a}$ is a suffix of $l_a$,}
\\
                            & \subword l_b \, u_{b,N} 
                &&\text{since $l_a\subword l_b$ by assumption,}
\\
                            & \subword v_{b,N} 
                &&\text{by definition of $l_b$,}
\\
                            & \subword v_{i,a} \, v_{b,N} = v(\tau)
\:.
\end{xalignat*}

2.\
Otherwise, $u_{i,a} = x \, l_a$ for some $x$, as illustrated in
Fig.~\ref{fig-cut1} where slanted arrows follow the rightmost embedding of
$u(\sigma)$ into $v(\sigma)$.
%
%%% File: fig-cut1.tex -*-LaTeX-*-
%%%
%
\begin{figure}[htbp]
\centering
{\setlength{\unitlength}{1.0mm} % relevant for width computation with \ubracew{}
\begin{gpicture}(107,33)(0,0)
%\put(0,0){\framebox(107,33){}}
\small
\gasset{AHnb=0,Nframe=n}
\drawline[dash={1 1},AHnb=1](7,12)(12,18)\node(h)(17,20){(rightmost embedding)}
\drawline[dash={1 1},AHnb=1](27,12)(29,18)
\drawline[dash={1 1},AHnb=1](41,12)(47.5,22)
\drawline[dash={1 1},AHnb=1](47,12)(56,18)
\drawline[dash={1 1},AHnb=1](67,12)(77.5,22)
\drawline[dash={1 1},AHnb=1](77,12)(80,18)
\drawline[dash={1 1},AHnb=1](107,12)(107,24)
{\gasset{linewidth=0.14,Nw=0.14,Nh=3,Nframe=y,ExtNL=y,NLangle=-90,NLdist=0.8}
\node(alabel)(7,2){$0$}
\node(alabel)(27,2){$i$}
\node(alabel)(47,2){$a$}
\node(alabel)(77,2){$b$}
\node(alabel)(107,2){$N$}
}
\gasset{Nframe=y,Nh=4,Nmr=0,ExtNL=y,NLangle=-90,NLdist=0.3}
\node[Nframe=n,Nw=0,NLangle=180,NLdist=1.5](u)(7,10){$u(\sigma)$:}
\node[Nw=20](ui)(17,10){$\ubracew{\hspace{19mm}}{u_{0,i}}$}
\node[Nw=20](ua)(37,10){$\ubracew{\hspace{19mm}}{u_{i,a}}$}
\node[Nw=30](ub)(62,10){$\ubracew{\hspace{29mm}}{u_{a,b}}$}
\node[Nw=30](uc)(92,10){$\ubracew{\hspace{29mm}}{u_{b,N}}$}
\gasset{NLangle=90}
\node[Nframe=n,Nw=0,NLangle=180,NLdist=1.5](u)(7,26){$v(\sigma)$:}
\node[Nw=20](vi)(17,26){$\obracew{\hspace{19mm}}{v_{0,i}}$}
\node[Nw=20](va)(37,26){$\obracew{\hspace{19mm}}{v_{i,a}}$}
\node[Nw=30](vb)(62,26){$\obracew{\hspace{29mm}}{v_{a,b}}$}
\node[Nw=30](vc)(92,26){$\obracew{\hspace{29mm}}{v_{b,N}}$}
\gasset{ExtNL=n,NLdist=0}
\node[Nw=14](ra)(34,10){$x$}
\node[Nw=6](ra)(44,10){$l_a$}
\node[Nw=10](rb)(72,10){$l_b$}
\end{gpicture}
}
\caption{Schematics for Lemma~\ref{cutblue}, with $l_a\subword l_b$}
\label{fig-cut1}
\end{figure}
%
%%% Local Variables:
%%% fill-column: 130
%%% ispell-check-comments: nil
%%% Local IspellDict: "english"
%%% End:

%
Here $u_{i,N} \subword v_{i,N}$
rewrites as $x \, l_a \, u_{a,N} \subword v_{i,a} \, v_{a,N}$. Now,
and since $l_a$ is (by definition) the longest suffix for which $l_a \, u_{a,N}\subword
v_{a,N}$, Lemma~\ref{basic}.{b} entails $x \subword v_{i,a}$. Then
\begin{xalignat*}{2}
u(\tau) = u_{i,a} \, u_{b,N} & = x \, l_a \, u_{b,N} 
\\
                            & \subword v_{i,a} \, l_b \, u_{b,N} 
                &&\text{since $x\subword v_{i,a}$ and $l_a\subword l_b$,}
\\
                            & \subword v_{i,a} \, v_{b,N} = v(\tau)
                &&\text{by definition of $l_b$.}
\end{xalignat*}

We can now infer  $u(\tau)\subword v(\tau)$ for any suffix $\tau \in R'$ (or for
$\tau=\sigma'$) from the corresponding $u(S(\tau))\subword v(S(\tau))$.
This shows that $\sigma'$ is a solution.
\qed
\end{proof}

If $i$ is a red index, i.e., if $u_{i,N}\not\subword v_{i,N}$, let
$r_i\in\Gamma^*$ be the shortest prefix of
$u_{i,N}$ such that $r_i^{-1}u_{i,N} \subword v_{i,N}$ (equivalently
$u_{i,N} \subword r_i \, v_{i,N}$) and call it the \emph{right margin} at $i$.
\begin{lem}[Cutting lemma for red indices]
\label{cutred}
Let $a < b$ be two congruent and red indices. If $r_b \subword r_a$, then
$\sigma' = \sigma[0,a)\sigma[b,N)$ is a solution (shorter than $\sigma$).
\end{lem}
\begin{proof}
Write $x$ for $r_b^{-1} u_{b,N}$. Then $u_{b,N}=r_b \, x$ and $x\subword v_{b,N}$. We
proceed as for Lemma~\ref{cutblue} and show that $u(S(\tau)) \subword
v(S(\tau))$ implies $u(\tau) \subword v(\tau)$ for all suffixes $\tau$ of
$\sigma'$. 
Assume $u(S(\tau)) \subword v(S(\tau))$ for some $\tau$. The
only interesting case is when $\tau \neq S(\tau)$, i.e., when $\tau =
\sigma[i,a)\sigma[b,N)$ for some $i<a$ (see Fig.~\ref{fig-cut2}).
%
%%% File: fig-cut2.tex -*-LaTeX-*-
%%%
%
\begin{figure}[htbp]
\centering
{\setlength{\unitlength}{1.0mm} % relevant for width computation with \ubracew{}
\begin{gpicture}(107,33)(0,0)
%\put(0,0){\framebox(107,33){}}
\small
\gasset{AHnb=0,Nframe=n}
\drawline[dash={1 1},AHnb=1](7,12)(12,18)\node(h)(17,20){(rightmost embedding)}
\drawline[dash={1 1},AHnb=1](27,12)(29,18)
\drawline[dash={1 1},AHnb=1](47,12)(42,20)
\drawline[dash={1 1},AHnb=1](57,12)(48.5,22)
\drawline[dash={1 1},AHnb=1](77,12)(70,20)
\drawline[dash={1 1},AHnb=1](83,12)(78,22)
\drawline[dash={1 1},AHnb=1](107,12)(107,24)
{\gasset{linewidth=0.14,Nw=0.14,Nh=3,Nframe=y,ExtNL=y,NLangle=-90,NLdist=0.8}
\node(alabel)(7,2){$0$}
\node(alabel)(27,2){$i$}
\node(alabel)(47,2){$a$}
\node(alabel)(77,2){$b$}
\node(alabel)(107,2){$N$}
}
\gasset{Nframe=y,Nh=4,Nmr=0,ExtNL=y,NLangle=-90,NLdist=0.3}
\node[Nframe=n,Nw=0,NLangle=180,NLdist=1.5](u)(7,10){$u(\sigma)$:}
\node[Nw=20](ui)(17,10){$\ubracew{\hspace{19mm}}{u_{0,i}}$}
\node[Nw=20](ua)(37,10){$\ubracew{\hspace{19mm}}{u_{i,a}}$}
\node[Nw=30](ub)(62,10){$\ubracew{\hspace{29mm}}{u_{a,b}}$}
\node[Nw=30](uc)(92,10){$\ubracew{\hspace{29mm}}{u_{b,N}}$}
\gasset{NLangle=90}
\node[Nframe=n,Nw=0,NLangle=180,NLdist=1.5](u)(7,26){$v(\sigma)$:}
\node[Nw=20](vi)(17,26){$\obracew{\hspace{19mm}}{v_{0,i}}$}
\node[Nw=20](va)(37,26){$\obracew{\hspace{19mm}}{v_{i,a}}$}
\node[Nw=30](vb)(62,26){$\obracew{\hspace{29mm}}{v_{a,b}}$}
\node[Nw=30](vc)(92,26){$\obracew{\hspace{29mm}}{v_{b,N}}$}
\gasset{ExtNL=n,NLdist=0}
\node[Nw=10](ra)(52,10){$r_a$}
\node[Nw=6](rb)(80,10){$r_b$}
\node[Nw=24](rx)(95,10){$x$}
\end{gpicture}
}
\caption{Schematics for Lemma~\ref{cutred}, with $r_b\subword r_a$}
\label{fig-cut2}
\end{figure}
%
%%% Local Variables:
%%% fill-column: 130
%%% ispell-check-comments: nil
%%% Local IspellDict: "english"
%%% End:

From $u_{i,N}=u_{i,a} \, u_{a,N}\subword v_{i,a} \, v_{a,N} = v_{i,N}$, i.e.,
$u(S(\tau)) \subword v(S(\tau))$, and $u_{a,N} \not \subword v_{a,N}$ (since
$a$ is a red index), Lemma~\ref{basic}.{c} entails $u_{i,a} \, r_a \subword
v_{i,a}$ by definition of $r_a$. Then
\begin{xalignat*}{2}
u(\tau) = u_{i,a} \, u_{b,N} = u_{i,a} \, r_b \, x & \subword
u_{i,a} \, r_a \, v_{b,N} 
&&\text{since $r_b\subword r_a$ and $x\subword v_{b,N}$,}
\\
 & \subword v_{i,a} \, v_{b,N} = v(\tau)
&&\text{since $u_{i,a}\,r_a\subword v_{i,a}$.}
\tag*{\qed}% since we don't have \qedhere
\end{xalignat*}
\end{proof}

For the next step let $g_1<g_2<\cdots<
g_{N_1}$ be all the blue indices in $\sigma$, and let $b_1<b_2<\cdots< b_{N_2}$
be the red indices. Observe that $N_1+N_2=N+1$ since each index in
$0,\ldots,N$ is either blue or red. We consider the corresponding sequences
$(l_{g_i})_{i=1,\ldots,N_1}$ of left margins and $(r_{b_i})_{i=1,\ldots,N_2}$ of right
margins.
\begin{lem}
\label{lem-blue&red-control}
$\size{l_{g_i}}\leq
(i-1)\times K_u$ for all $i=1,\ldots,N_1$, and $\size{r_{b_i}}\leq (N_2-i +
1)\times K_u$ for all $i=1,\ldots,N_2$.
In other words, the sequence of left margins and the \emph{reversed}
sequence of right margins are $K_u$-controlled.
\end{lem}
\begin{proof}
We prove that $\size{l_{g_i}}\leq(i-1)\times K_u$ by induction on $i$, showing
$\size{l_{g_1}}=0$ and $\size{l_{g_i}}-\size{l_{g_{i-1}}}\leq K_u$ for
  $i>1$.

The base case $i=1$ is easy: obviously $g_1=0$ since $0$ is a blue index, and
$l_0=\eps$ since it is the only suffix of $u_{0,0}=\eps$, so that
$\size{l_{g_1}}=0$.

For the inductive step $i>1$, write $p$ for $g_{i-1}$ and $q$ for $g_i$. By
definition, $l_p$ is the longest suffix of $u_{0,p}$ with $l_p\,u_{p,N}=
l_p\, u_{p,q} \, u_{q,N}\subword v_{p,N}$. Since $l_q \, u_{q,N}\subword
v_{q,N}\subword v_{p,N}$, $l_q$ must be a suffix of $l_p\, u_{p,q}$, hence
$\size{l_q}\leq \size{l_p}+ \size{u_{p,q}}\leq \size{l_p}+K_u(q-p)$. This
proves the claim in the case where $q=p+1$, i.e., when $p$ and $p+1$ are
blue.

There remains the case where $q>p+1$ and where all the indices from
$p+1$ to $q-1$ are red. Thus in particular $u_{q-1,N}=u_{q-1,q}\,
u_{q,N}\not\subword v_{q-1,N}$. On the other hand $q$ is blue and 
 $l_q \, u_{q,N}\subword v_{q,N}\subword v_{q-1,N}$. We conclude that
$l_q$ must be a suffix of $u_{q-1,q}$, so that $\size{l_q}\leq K_u$ which
proves the claim.
\\

The reasoning for $\size{r_{b_i}}$ is similar:

If $b_{i+1}=b_i+1$, then both $b_i$ and the next index are red. Then
$r_{b_i}$ is a prefix of $u_{b_i,b_i+1}\,r_{b_{i+1}}$ so that
$\size{r_{b_i}}\leq K_u+\size{r_{b_{i+1}}}$.

If $b_{i+1}>b_i+1$, then $b_{i}+1$ is blue and $r_{b_i}$ is a prefix of
$u_{b_i,b_i+1}$ so that $\size{r_{b_i}}\leq K_u$. 

For the base case, we have  $b_{N_2}<N$ since
$N$ is blue. Hence $b_{N_2}+1$ is blue and
$\size{r_{b_{N_2}}}\leq K_u$ as above.

Finally, $\size{r_{b_i}}\leq (N_2+1-i)\times K_u$ for all $i=1,\ldots,N_2$. 
\qed
\end{proof}

We are now ready to conclude the proof of Theorem~\ref{theo-short}.
Let $N_c\egdef \mu(R) \, \mu(R')+1$ and $L\egdef H(N_c, K_u, \Gamma)$
and assume that $N > 2L$. Since $N_1+N_2=N+1$,
either $\sigma$ has at least 
$L+1$ blue indices and, by definition of $L$ and $H$, there exist $N_c$
blue indices $a_1<a_2<\cdots<a_{N_c}$ with $l_{a_1}\subword l_{a_2}\subword
\cdots\subword l_{a_{N_c}}$, or $\sigma$ has at least $L+1$ red indices and
there exist $N_c$ red indices $a'_1<a'_2<\cdots<a'_{N_c}$ with
$r_{a'_{N_c}}\subword \cdots\subword r_{a'_2}\subword r_{a'_1}$ (since it
is the reversed sequence of right margins that is controlled).
Out of $N_c= \mu(R) \, \mu(R') + 1$ indices,   two
 must be congruent, fulfilling the assumptions of
either Lemma~\ref{cutblue} or Lemma~\ref{cutred}. Therefore $\sigma$ can be cut to
obtain a shorter solution.
\\

Since $\PEPpd$ and $\PEPpcod$ are equivalent problems modulo mirroring of
$R$, $u$ and $v$, we deduce that  $\PEPpd$ too is decidable, and more precisely:
\begin{cor}
\label{coro-short}
A $\PEPpd$ instance has a solution if, and only if, it has a
solution of length at most $2L$. 
\end{cor}

%%% Local Variables:
%%% fill-column: 75
%%% ispell-check-comments: nil
%%% Local IspellDict: "english"
%%% End:
% LocalWords:

% LocalWords:  codirect fsttcs AHnb Nframe

%%% File: sec-infpeprpd.tex -*-LaTeX-*-

\section{Counting the number of solutions}
%=========================================
\label{sec-infpeprpd}

We consider two counting questions: $\EXINF\PEPpd$ is the question
whether a $\PEPpd$ instance has infinitely many solutions (a decision
problem), while $\COUNT\PEPpd$ is the problem of computing the number of
solutions of the instance (a number in $\Nat\cup\{\infty\}$). For technical
convenience, we often deal with the (equivalent) codirected versions,
$\EXINF\PEPpcod$
and $\COUNT\PEPpcod$.

For an instance $(\Sigma, \Gamma, u,v,R,R')$, we let
$K_v\egdef\max_{a\in\Sigma}\size{v(a)}$ and define
\begin{xalignat*}{2}
M & \egdef H(\mu(R) \, \mu(R') +1, K_v, {\Gamma}) \:,
&
M' & \egdef H\bigl((2M+2)\mu(R) \, \mu(R') + 1, K_u, {\Gamma}\bigr) \:.
\end{xalignat*}
In this section we prove:
\begin{thm}
\label{theo-long}
For a $\PEPpd$ or $\PEPpcod$ instance, the following are equivalent:
\begin{description}
\item[(a)] it has infinitely many solutions;
\item[(b)] it has solution of length $N$ with $2M<N$;
\item[(c)] it has a solution of length $N$ with $2M<N\leq 2M'$.
\end{description}
\noindent
This entails the decidability of $\EXINF\PEPpd$ and $\EXINF\PEPpcod$, and
the computability of
$\COUNT\PEPpd$ and $\COUNT\PEPpcod$.
\end{thm}
As with Theorem~\ref{theo-short}, the length bounds $2M$ and $2M'$ are
computable, so that  $\EXINF\PEPpd$ and $\EXINF\PEPpcod$ can be decided by
finite enumeration. When the number of solutions is finite, 
counting them can also be done by finite enumeration
since we know all solutions have then length at most $2M$. 

For the proof of Theorem~\ref{theo-long}, we  
first observe that if the instance has a  solution of length $N>2M$, it has a solution with $R$
replaced by $R^{>}\egdef R \cap \Sigma^{2M+1}\Sigma^*$. The syntactic congruence
associated with $R^{>}$ has index at most
$(2M+2)\mu(R)$. From Theorem~\ref{theo-short}, we deduce that
the modified
instance has a solution of length at most $2M'$.
Hence (b) and (c) are equivalent.

It remains to  show that (b) implies (a)
since obviously (a) implies (b).
For this we fix an arbitrary $\PEPpcod$
instance $(\Sigma, \Gamma, u, v, R, R')$ and consider a solution $\sigma$,
of length $N$. We develop two so-called ``iteration lemmas'' that are
similar to the cutting lemmas from Section~\ref{sec-peprpd}, with the
difference that they expand $\sigma$ instead of reducing it.

As before, an index $i \in [0,N]$ is said to be \emph{blue} if $u_{i,N}
\subword v_{i,N}$, and \emph{red} otherwise. With a blue (resp., a red)
index $i\in[0,N]$ we associate a word $s_i$ (resp., $t_i$) in $\Gamma^*$.
The $s_i$'s and $t_i$'s are analogous to the $l_i$'s and $r_i$'s from
Section~\ref{sec-peprpd}, however they are factors of $v(\sigma)$, not
of $u(\sigma)$ like $l_i$ or $r_i$, and this explains the difference between
$M$ and $L$. The terms ``left margin'' and ``right margin'' will be reused
here for these factors.
\\

We start with blue indices. For a blue index $i\in[0,N]$, let
$s_i$ be the longest prefix of $v_{i,N}$ such that $u_{i,N}
\subword s_i^{-1}v_{i,N}$ (equivalently, such that $s_i \, u_{i,N} \subword v_{i,N}$) and
call it the \emph{right margin} at $i$.
\begin{lem}
\label{iterationblueab}
Suppose $a < b$ are two blue indices with $s_b \subword s_a$. Then for all
$k \geq 1$, $s_a (u_{a,b})^k \subword (v_{a,b})^k s_b$.
\end{lem}
\begin{proof}
$s_a \, u_{a,N}\subword v_{a,N}$ expands as $(s_a \, u_{a,b})
  u_{b,N}\subword v_{a,b} \, v_{b,N}$. Since $b$ is blue, $u_{b,N}\subword
  v_{b,N}$ and, by definition of $s_b$, Lemma~\ref{basic}.{d}
  further yields $s_a \, u_{a,b}\subword v_{a,b} \, s_b$.
One concludes with Lemma~\ref{basic}.{f}, using
 $s_b\subword s_a$.
\qed
\end{proof}

\begin{lem}[Iteration lemma for blue indices]
\label{pumpblue}
Let $a < b$ be two congruent  blue indices. If $s_b \subword s_a$, then
for every $k\geq 1$, $\sigma'= \sigma[0,a).\sigma[a,b)^k.\sigma[b,N)$ is a
      solution.
\end{lem}
\begin{proof}
Let $\tau$ be any suffix of $\sigma'$. We show that $u(\tau) \subword
v(\tau)$ when $\tau \in R'$ or $\tau = \sigma'$, which will complete the
proof. There are three cases, depending on how long $\tau$ is.
\begin{itemize}
\item
$\tau$ is a suffix of $\sigma[a,N)$. Then $\tau$ is a suffix of $\sigma$
  itself, and this case is trivial since $\sigma$ is a solution.
\item
$\tau$ is $\sigma[i,b)\sigma[a,b)^p\sigma[b,N)$ for some $p \geq 1$ and $a
      < i \leq b$. Since $a$ and $b$ are congruent, $\tau \in R'$ implies
      $\sigma[i,N) \in R'$. Thus $u_{i,N} \subword v_{i,N}$, hence
        $u_{i,b}\subword v_{i,b}\, s_b$ (since $u_{b,N}\subword v_{b,N}$). 
\begin{align*}
u(\tau) &= {u_{i,b}} (u_{a,b})^p \, u_{b,N} \\
 &\subword v_{i,b} \,{s_b} (u_{a,b})^p \, u_{b,N} \\
 &\subword v_{i,b} \, s_a (u_{a,b})^p \,u_{b,N} && \text{since } s_b \subword s_a\\
 &\subword v_{i,b} (v_{a,b})^p \, s_b \, u_{b,N} && \text{by Lemma~\ref{iterationblueab}}\\
 &\subword v_{i,b} (v_{a,b})^p \, v_{b,N} && \text{by definition of } s_b \\
 &=v(\tau) 
\:.
\end{align*}
\item
$\tau$ is $\sigma[i,a)\sigma[a,b)^k\sigma[b,N)$ for some $0 \leq i < a$.
      Since $a$ and $b$ are congruent, $\tau\in R'$ (or $\tau=\sigma$)
      implies $u_{i,N}\in R'$ (or $u_{i,N}=\sigma$) so that $u_{i,N}
      \subword v_{i,N}$, from which we deduce $u_{i,a}\subword v_{i,a} \, s_a$
      as in the previous case. Then, using Lemma~\ref{iterationblueab} and
      $s_b \, u_{b,N}\subword v_{b,N}$, we get
\begin{align*}
\notag
u(\tau) &= {u_{i,a}}(u_{a,b})^{k} \, u_{b,N} \\
&\subword v_{i,a} \, s_a (u_{a,b})^{k} \, u_{b,N} \\
&\subword v_{i,a} (v_{a,b})^{k} \, s_b \, u_{b,N} && \text{by Lemma~\ref{iterationblueab}}  \\
&\subword v_{i,a} (v_{a,b})^{k} \, v_{b,N} && \text{by definition of } s_b\\
&=v(\tau)
\: .
\tag*{\qed}% since we don't have \qedhere
\end{align*}
\end{itemize}
\end{proof}

Now to red indices. For a red index $i\in [0,N]$,
let $t_i$ be the shortest suffix of $v_{0,i}$
such that $u_{i,N} \subword t_i \, v_{i,N}$. This is called the \emph{left margin} at $i$.
Thus, for a blue $j$ such that $j<i$,
$u_{j,N}\subword v_{j,N}$ implies $u_{j,i} \, t_i\subword v_{j,i}$ by
% Should we point out that $t_i$ is a suffix of $v_{j,i}$?
% This is like the lemma sasb above. We can probably also leave it unstated.
% -Prateek
Lemma~\ref{basic}.{e}.
\begin{lem}[Iteration lemma for red indices]
\label{pumpred}
Let $a < b$ be two congruent  red indices. If $t_a \subword t_b$, then
for every $k \geq 1$, $\sigma'=\sigma[0,a).\sigma[a,b)^k.\sigma[b,N)$ is a solution.
\end{lem}

\begin{proof}
Let $\tau$ be any suffix of $\sigma'$. We show that $u(\tau)\subword
v(\tau)$ when $\tau\in R'$ or $\tau=\sigma'$, which will complete the
proof. There are three cases, depending on how long $\tau$ is.
\begin{itemize}
\item
$\tau$ is a suffix of $\sigma[a,N)$. Then $\tau$ is a suffix of $\sigma$
  itself, and this case is trivial since $\sigma$ is a solution.
\item
$\tau$ is $\sigma[i,b)\sigma[a,b)^p\sigma[b,N)$ for some $p \geq 1$ and $a
      < i \leq b$. Since $a$ and $b$ are congruent, $\tau \in R'$ implies
      $\sigma[i,N) \in R'$ and so $u_{i,N} \subword v_{i,N}$. By definition of
        $t_a$, we have
        $u_{a,b}u_{b,N} \subword (t_a v_{a,b})v_{b,N}$. Using
        Lemma~\ref{basic}.{e} and the definition of
        $t_b$ we get $u_{a,b} \,
        t_b \subword t_a \, v_{a,b}$,
        and then $(u_{a,b})^p \, t_b \subword t_a (v_{a,b})^p$ with
        Lemma~\ref{basic}.{g}. Then
\begin{align*}
u(\tau) &= u_{i,b} (u_{a,b})^p \, {u_{b,N}} \\
&\subword u_{i,b} (u_{a,b})^p \, t_b \, v_{b,N} && \text{by definition of } t_b\\
&\subword u_{i,b} \, t_a (v_{a,b})^p \, v_{b,N} && \text{as above}\\
&\subword u_{i,b} \, t_b (v_{a,b})^p \, v_{b,N} && \text{since } t_a \subword t_b\\
&\subword v_{i,b} (v_{a,b})^p \, v_{b,N} && \text{since $u_{i,N} \subword v_{i,N}$, $b$ is red, Lemma~\ref{basic}.{e}} \\
&=v(\tau)
\: .
\end{align*}
\item
$\tau$ is $\sigma[i,a)\sigma[a,b)^k\sigma[b,N)$ for some $0 \leq i < a$ and
      $k\geq 1$.
Since $a$ and $b$ are congruent, $\tau\in R'$ (or $\tau=\sigma$) implies
$u_{i,N}\in R'$ (or $u_{i,N}=\sigma$) so that $u_{i,N}\subword v_{i,N}$,
from which we deduce $u_{i,a} \, t_a\subword v_{i,a}$ as in the previous case. Then
\begin{align*}
u(\tau) &= u_{i,a} (u_{a,b})^{k} \, u_{b,N} \\
& \subword u_{i,a} (u_{a,b})^{k} \, t_b \, v_{b,N} && \text{by definition of } t_b \\
& \subword u_{i,a} \, t_a \, (v_{a,b})^{k} \, v_{b,N} && \text{as before} \\
& \subword v_{i,a} (v_{a,b})^{k} \, v_{b,N} && \text{as above}\\
& = v(\tau)
\: .
\tag*{\qed} % since we don't have \qedhere
\end{align*}
\end{itemize}
\end{proof}

We may now prove that the $\PEPpcod$ instance has infinitely many solutions
if it has solution of length $N>2M$, i.e., that (b) implies (a) in
Theorem~\ref{theo-long}.

Suppose there are $N_1$ blue indices in $\sigma$, say $g_1<g_2<\cdots<
g_{N_1}$; and $N_2$ red indices, say $b_1<b_2<\cdots<
b_{N_2}$.
\begin{lem}
\label{lem-blue&red-control2}
$\size{s_{g_i}}\leq
(N_1 - i+1)\times K_v$ for all $i=1,\ldots,N_1$, and $\size{t_{b_i}}\leq
(i-1)\times K_v$ for all $i=1,\ldots,N_2$.
That is, the \emph{reversed} sequence of right margins and the sequence of
left margins are $K_v$-controlled.
\end{lem}

\begin{proof}
We start with blue indices and right margins.
\begin{lem}
\label{sasb}
Suppose $a < b$ are two blue indices. Then $s_a$ is a prefix of $v_{a,b} \,
s_b$.
\end{lem}
\begin{proof}
Both $s_a$ and $v_{a,b}s_b$ are prefixes of $v_{a,N}$, hence one of them is
a prefix of the other. Assume, by way of contradiction, that $v_{a,b} \, s_b$
is a proper prefix of $s_a$, say $s_a = v_{a,b} \, s_b \, x$ for some $x \neq
\epsilon$. Then $s_a \, u_{a,N}\subword v_{a,N}$ rewrites as $v_{a,b} \,
s_b \, x \,
u_{a,N}\subword v_{a,b} \, v_{b,N}$. Cancelling $v_{a,b}$ on both sides gives
$s_b \, x \, u_{a,N}\subword v_{b,N}$, i.e., $(s_b \, x \, u_{a,b})u_{b,N} \subword v_{b,N}$,
which contradicts the definition of $s_b$.
\qed
\end{proof}

We now show that $s_{g_{N_1}}, \ldots, s_{g_1}$ is $K_v$-controlled.
$N$ is a blue index, and $\size{s_N} = 0$. For $i \in
[0,N)$, if both $i$ and $i+1$ are blue indices, then by Lemma~\ref{sasb},
$\size{s_i} \leq \size{s_{i+1}} + K_v$. If $i$ is blue and $i+1$ is red,
then it is easy to see that $s_i$ is a prefix of $v(\sigma_i)$, and hence
$\size{s_i} \leq K_v$. So we get that $s_{g_{N_1}}, \ldots, s_{g_1}$
is $K_v$-controlled.
\\

Now to red indices and left margins.
$0$ is not a red index. For $i \in [0,N)$, if both $i$
and $i+1$ are red, then it is easy to see that $t_{i+1}$ is a suffix of
$t_i \, v(\sigma_i)$, and so $\size{t_{i+1}} \leq \size{t_i} + K_v$. If $i$
is blue and $i+1$ is red, then $t_{i+1}$ is a suffix of $v(\sigma_i)$,
and so $\size{t_{i+1}} \leq K_v$.  So we get that $t_{b_1}, \ldots, t_{b_{N_2}}$ is
$K_v$-controlled.
\qed
\end{proof}

Assume that $\sigma$ is a long solution of length $N > 2M$. At least
$M+1$ indices among $[0,N]$ are blue, or at least $M+1$ are red. We apply
one of the two above claims, and from either $s_{g_{N_1}}, \ldots, s_{g_1}$
(if $N_1 > M$) or $t_{b_1}, \ldots, t_{b_{N_2}}$ (if $N_2 > M$) we
get an increasing subsequence of length $\mu(R)\,\mu(R')+1$. Among these
there must be two congruent indices. Then we get infinitely many solutions
by Lemma~\ref{pumpblue} or Lemma~\ref{pumpred}.

\iffalse % skip this !!!
\begin{remark}
The above proof extends the proof
in~\cite{CS-countpep} for $\EXINF\PEP$. Extending to
$\EXINF\PEPpcod$ raises two issues. First, the iteration lemma has
to handle codirectness, i.e., suffixes in $R'$. Second, two distinct lemmas
are needed, for blue and for red indices, while only one was needed for
$\EXINF\PEP$ because of mirror properties of the problem.
%
\qed
\end{remark}
\fi

%%% Local Variables:
%%% fill-column: 75
%%% ispell-check-comments: nil
%%% Local IspellDict: "english"
%%% End:
% LocalWords:

% LocalWords:  codirected codirectness Theo fsttcs

%%% File: sec-universal.tex -*-LaTeX-*-

\section{Universal variants of $\PEPpd$}
%=======================================
\label{sec-universal}

We consider universal variants of $\PEPpd$ (or rather $\PEPpcod$ for the
sake of uniformity). Formally, given instances
$(\Sigma,\Gamma,u,v,R,R')$ as usual, $\ALL\PEPpcod$ is the question
whether \emph{every} $\sigma\in R$ is a solution, i.e., satisfies both
$u(\sigma)\subword v(\sigma)$ and $u(\tau)\subword v(\tau)$ for all
suffixes $\tau$ that belong to $R'$. Similarly, $\ALLINF\PEPpcod$
is the question whether ``almost all'', i.e., \emph{all but finitely many}, $\sigma$ in $R$ are solutions,
and $\COUNT\neg\PEPpcod$ is the associated counting problem that asks how
many $\sigma\in R$ are not solutions.

These universal questions can also be seen as Post \emph{non-}embedding
problems, asking whether there exists some $\sigma\in R$ such that
$u(\sigma)\not\subword v(\sigma)$? Introduced in~\cite{CS-countpep} with
$\ALL\PEP$, they are significantly less challenging than the standard
$\PEP$ problems, and decidability is easier to establish. For this reason,
we just show in this article how $\ALL\PEPpcod$ and $\ALLINF\PEPpcod$
reduce to $\ALLINF\PEP$ whose decidability was shown in~\cite{CS-countpep}.
The point is that partial codirectness constraints can be eliminated since
universal quantifications commute with conjunctions (and since the
codirectness constraint is universal itself).

\begin{lem}
\label{lem-univ-pcod-reduce}
$\ALL\PEPpcod$ and $\ALLINF\PEPpcod$ many-one reduce to
  $\ALLINF\PEP$.
\end{lem}
\begin{cor}
$\ALL\PEPpcod$ and $\ALLINF\PEPpcod$ are decidable,
  $\COUNT\neg\PEPpcod$ is computable.
\end{cor}

%reduction from \ALL\PEPpcod to \ALL\PEP. Not needed anymore
\iffalse
A $\ALL\PEPpcod$ instance with $R,R'$ as above is negative iff there is
a $\sigma\in R$ with $u(\sigma)\not\subword v(\sigma)$ or with a suffix
$\tau\in R'$ such that $u(\tau)\not\subword v(\tau)$.
Let now $\overrightarrow{R}$ be the set of all suffixes of words in $R$,
that is, $\overrightarrow{R} \egdef \{x ~|~ \exists y (yx \in R)\}$. Then
the $\ALL\PEPpcod$ instance is negative iff the $\ALL\PEP$ instance
over $R_2\egdef R \cup (\overrightarrow{R} \cap R')$ is negative. Since
$\overrightarrow{R}$ and $R_2$ are regular and computable from
$R$ and $R'$, the reduction is effective.
\qed
\fi

We now prove Lemma~\ref{lem-univ-pcod-reduce}.
First, $\ALL\PEPpcod$ easily reduces to
$\ALLINF\PEPpcod$:
%\footnote{$\ALL\PEPpcod$ also many-one reduces to $\ALL\PEP$ but this is not so useful for the second half of the Lemma.}
add an extra letter $\ttz$ to $\Sigma$ with
$u(\ttz)=v(\ttz)=\epsilon$ and replace $R$ and $R'$ with
$R.\ttz^*$ and $R'.\ttz^*$. Hence the
second half of the lemma entails its first half by transitivity of
reductions.

For reducing $\ALLINF\PEPpcod$, it is easier to start
with the negation of our question:
\begin{gather}
\label{eq-neg-question}
\tag{$\ast$}
\exists^\infty \sigma \in R : \bigl( u(\sigma) \not \subword v(\sigma)
\textrm{ or $\sigma$ has a suffix $\tau$ in $R'$ with } u(\tau) \not \subword v(\tau)\bigr)
\:.
\end{gather}
Call $\sigma \in R$ a \emph{type 1 witness} if $u(\sigma) \not
\subword v(\sigma)$, and a \emph{type 2 witness} if it has a suffix
$\tau \in R'$ with $u(\tau) \not \subword v(\tau)$.
Statement~\eqref{eq-neg-question} holds if, and only if, there are infinitely many type
1 witnesses or  infinitely many type 2 witnesses.
The existence of infinitely many type
1 witnesses (call that ``case 1'') is the negation of a $\ALLINF\PEP$ question.
Now suppose that there are infinitely many type 2
witnesses, say $\sigma_1, \sigma_2, \ldots$ For each $i$, pick a
suffix $\tau_i$ of $\sigma_i$ such that $\tau_i \in R'$ and $u(\tau_i) \not
\subword v(\tau_i)$. The set $\{\tau_i ~|~ i=1,2,\ldots\}$ of these
suffixes can be finite or infinite. If it is infinite (``case 2a''), then
\begin{gather}
\label{the-t2-prop}
\tag{$\ast\ast$}
u(\tau) \not \subword v(\tau)
\text{ for infinitely many }
\tau \in (\overrightarrow{R} \cap R')
\:,
\end{gather}
where $\overrightarrow{R}$ is short for $\overrightarrow{{}^{\geq 0}R}$ and
for $k\in\Nat$, $\overrightarrow{{}^{\geq k}R} \egdef \{y ~|~ \exists x : (
\size{x} \geq k \textrm{ and } xy \in R)\}$ is the set of the suffixes of
words from $R$ one obtains by removing at least $k$ letters. Observe that,
conversely, \eqref{the-t2-prop} implies the existence of infinitely many
type 2 witnesses (for a proof, pick $\tau_1 \in \overrightarrow{R} \cap R'$
satisfying the above, choose $\sigma_1 \in R$ of which $\tau_1$ is a
suffix. Then choose $\tau_2$ such that $\size{\tau_2} > \size{\sigma_1}$,
and proceed similarly).

On the other hand, if $\{\tau_i ~|~ i=1,2,\ldots\}$ is finite (``case 2b''), then there
is a $\tau \in R'$ such that $u(\tau) \not \subword v(\tau)$ and
$\sigma'\tau \in R$ for infinitely many $\sigma'$. By a standard pumping
argument, the second point is equivalent to the existence of some
such $\sigma'$ with also $\size{\sigma'} > k_R$, where
$k_R$ is the size of a NFA for $R$ (taking $k_R=\mu(R)$ also works). Write
now $\hat R$ for $\overrightarrow{{}^{>k_R}R}$:
if $\{\tau_i
~|~ i=1,2,\ldots\}$ is finite, then $u(\tau) \not\subword v(\tau)$ for
some $\tau$ in $(R' \cap \hat R)$, and conversely this implies the
existence of infinitely many type 2 witnesses.

To summarize, and since $\overrightarrow{R}$ and
$\hat R$ are regular and effectively computable from $R$,
we have just reduced $\ALLINF\PEPpcod$ to the
following conjunction
\begin{xalignat*}{2}
&\forall^\infty \sigma \in R : u(\sigma)  \subword v(\sigma)
&&\texttt{(not case 1)}
\\
\bigwedge\;
&\forall^\infty \tau \in (\overrightarrow{R} \cap R') : u(\tau) \subword
v(\tau)
&&\texttt{(not case 2a)}
\\
\bigwedge\;%was\textrm{ and }
&\forall \tau \in (\hat{R} \cap R') : u(\tau) \subword v(\tau)
\:.
&&\texttt{(not case 2b)}
\end{xalignat*}
This is now reduced to a single $\ALLINF\PEP$ instance by
rewriting the $\ALL\PEP$ into a $\ALLINF\PEP$ (as explained
in the beginning of this proof)
and relying on a distributivity property of the form
\[
\bigwedge_{i=1}^n
\Bigl[\forall^\infty
\sigma\in R_i : u(\sigma)\subword v(\sigma)
\Bigr]
\:\equiv\:
\forall^\infty
\sigma\in  \Bigl[ \bigcup_{i=1}^n R_i \Bigr] : u(\sigma)\subword v(\sigma)
\]
to handle the resulting conjunction of $3$ $\ALLINF\PEP$ instances.

%%% Local Variables:
%%% fill-column: 75
%%% ispell-check-comments: nil
%%% Local IspellDict: "english"
%%% End:

% LocalWords:  codirectness yx eq xy NFA

%%% File: sec-undec.tex -*-LaTeX-*-

\section{Undecidability for $\PEPdcd$ and other extensions}
%==========================================================
\label{sec-undec}

The decidability of $\PEPpd$ is a non-trivial generalization of previous
results for $\PEP$. It is a natural question whether one can
further generalize the idea of partial directness and maintain
decidability.
In this section we describe
two attempts that lead to undecidability, even though they remain inside
the regular $\PEP$ framework.\footnote{
  $\PEP$  is undecidable if we allow constraint sets $R$ outside
  $\Reg(\Sigma)$~\cite{CS-fsttcs07}.
   Other extensions, like $\exists x\in R_1:\forall y\in
  R_2:u(xy)\subword v(xy)$, for $R_1,R_2\in\Reg(\Sigma)$, have been shown
  undecidable~\cite{CS-dlt2010}.}

\paragraph{Allowing non-regular $R'$.}
%=====================================
One direction for extending $\PEPpd$ is to allow \emph{more expressive $R'$
  sets} for partial (co)directness. Let $\PEPdcflpcod$ and
$\PEPprespcod$ be like $\PEPpcod$ except that $R'$ can be any
deterministic context-free $R'\in\DCFL(\Sigma)$ (resp.,
any
Presburger-definable $R'\in\Pres(\Sigma)$, i.e., a language consisting
of all words whose Parikh image lies in a given
Presburger, or semilinear, subset of $\Nat^{\size{\Sigma}}$).
Note that $R\in\Reg(\Sigma)$ is still required.
\begin{thm}[Undecidability]
\label{theo-undec-dcfl-presburger}
$\PEPdcflpcod$ and $\PEPprespcod$ are $\Sigma_1^0$-complete.
\end{thm}
Since both problems clearly are in $\Sigma_1^0$, one only has to prove
hardness by reduction, e.g., from $\PCP$, Post's Correspondence Problem.
Let $(\Sigma, \Gamma, u,v)$ be a $\PCP$ instance (where the question
is whether  there exists $x \in \Sigma^+$ such that
$u(x) = v(x)$). Extend $\Sigma$ and
$\Gamma$ with new symbols: $\Sigma' \egdef \Sigma \cup \{1,2\}$ and
$\Gamma' \egdef \Gamma \cup \{\#\}$. Now define
$u',v':\Sigma'{}^*\rightarrow\Gamma'{}^*$ by extending $u,v$ on the new
symbols with $u'(1) = v'(2) = \epsilon$ and $u'(2) = v'(1) = \#$.
Define now $R = 12\Sigma^+$ and $R' = \{\tau 2\tau' ~|~ \tau,\tau' \in
\Sigma^* \text{ and } \size{u(\tau\tau')}\neq\size{v(\tau\tau')}\}$. Note that $R'$ is
deterministic context-free and Presburger-definable.
\begin{lem}
\label{lem-undec-dcfl-presburger}
The $\PCP$ instance $(\Sigma, \Gamma, u,v)$ has a solution if and only if
the $\PEPprespcod$ and $\PEPdcflpcod$ instance $(\Sigma', \Gamma', u', v',
R, R')$ has a solution.
\end{lem}

\begin{proof}
Suppose $\sigma$ is a solution to the $\PCP$ problem. Then $\sigma \neq
\epsilon$ and $u(\sigma) = v(\sigma)$. Now $\sigma'\egdef 12\sigma$ is a
solution to the partially codirected problem since $12\sigma\in R$,
$u'(12\sigma)=\# u(\sigma)\subword v'(12\sigma)=\# v(\sigma)$, and $\sigma'$
has
no suffix in $R'$ (indeed $2\sigma\not\in R'$ since
$\size{u(\sigma)}=\size{v(\sigma)}$).

Conversely, suppose $\sigma'$ is a solution to the partially codirected
problem. Then $\sigma' = 12\sigma$ for some $\sigma \neq \epsilon$. Since
$u'(\sigma')=\#u(\sigma) \subword v'(\sigma')=\#v(\sigma)$, we have
$u(\sigma) \subword v(\sigma)$. If $\size{u(\sigma)} \neq
\size{v(\sigma)}$, then $2\sigma \in R'$, and so we must have
$u'(2\sigma)=\#u(\sigma) \subword v'(2\sigma)=v(\sigma)$. This is not
possible as $\#$ does not occur in $v(\sigma)$. So $\size{u(\sigma)} =
\size{v(\sigma)}$, and $u(\sigma) = v(\sigma)$. Thus $\sigma$ is a
solution to the $\PCP$ problem. \qed
\end{proof}

\paragraph{Combining directness and codirectness.}
%=================================================
Another direction is to allow \emph{combining} directness and codirectness
constraints. Formally, $\PEPdcd$ is the problem of deciding, given
$\Sigma$, $\Gamma$, $u$, $v$, and $R\in\Reg(\Sigma)$ as usual, whether
there exists $\sigma \in R$ such that $u(\tau) \subword v(\tau)$ and
$u(\tau')\subword v(\tau')$ for all decompositions $\sigma=\tau.\tau'$. In
other words, $\sigma$ is both a direct and a codirect solution.

Note that $\PEPdcd$ has no $R'$ parameter (or, equivalently, has
$R'=\Sigma^*$) and requires directness and codirectness at all
positions. However, this restricted combination is already undecidable:
\begin{thm}[Undecidability]
\label{theo-undec-pepdcd}
$\PEPdcd$ is $\Sigma_1^0$-complete.
\end{thm}
Membership in $\Sigma_1^0$ is clear and we prove hardness by reducing from the
Reachability Problem for length-preserving semi-Thue systems.
%%
%% The undecidability
%% is linked to relying on \emph{different} embeddings of $u(\sigma)$ in $v(\sigma)$
%% for the directness and codirectness. In contrast, for $\PEPpd$ we need to consider
%% only the leftmost embedding of $u(\sigma)$ in $v(\sigma)$.
%%

A semi-Thue system $S=(\Upsilon,\Delta)$ has a finite set $\Delta\subseteq
\Upsilon^*\times\Upsilon^*$ of string rewrite rules over some finite alphabet
$\Upsilon$, written $\Delta = \{ l_1 \rightarrow r_1, \ldots,
l_k\rightarrow r_k\}$.
The one-step rewrite relation
 $\rew\subseteq\Upsilon^*\times\Upsilon^*$ is
defined as usual with $x\rew y$ $\equivdef$
$x=z l z'$ and
$y=z r z'$ for some rule $l\rightarrow r$ in $\Delta$ and strings $z,z'$ in
$\Upsilon^*$.
We write $x\rewn{m} y$ and $x\rews y$ when $x$ can be rewritten
into $y$ by a sequence of $m$ (respectively, any number, possibly zero)
rewrite steps.

The \emph{Reachability Problem} for semi-Thue systems is ``Given
$S=(\Upsilon,\Delta)$ and two regular languages $P_1,P_2\in\Reg(\Upsilon)$,
is there $x\in P_1$ and $y\in P_2$ s.t.\ $x\rews y$?''.
It is well-known (or easy to see by encoding Turing
machines in semi-Thue systems) that this problem is
undecidable (in fact, $\Sigma_1^0$-complete) even when
restricted to \emph{length-preserving systems}, i.e., systems where
$\size{l}=\size{r}$ for all rules $l\rightarrow r\in\Delta$.
\\

We now construct a many-one reduction to $\PEPdcd$.
Let $S=(\Upsilon,\Delta)$, $P_1$, $P_2$ be a length-preserving instance of
the Reachability Problem. W.l.o.g., we assume $\epsilon\not\in P_1$ and we
restrict to reachability via an even and non-zero number of rewrite
steps. With any such instance we associate a $\PEPdcd$ instance
$u,v:\Sigma^*\to\Gamma^*$ with $R\in\Reg(\Sigma)$ such that the
following Correctness Property holds:
\begin{gather}
\label{eq-reduction-spec}
\tag{CP}% was {$\ast\ast\ast$}
\begin{array}{rl}
& \exists x\in P_1,\: \exists y\in P_2,\: \exists m \text{ s.t.\ }
  x\rewn{m} y \text{ (and $m>0$ is even)}
\\[.5em]
\text{ iff } &
\exists \sigma\in R
\text{ s.t.\ }
 \sigma=\tau \tau' \text{ implies }
u(\tau)\subword v(\tau) \text{ and } u(\tau')\subword v(\tau')
\:.
\end{array}
\end{gather}
The reduction uses letters like $\tta$, $\ttb$ and $\ttc$ taken from $\Upsilon$, and
adds $\xs$ as an extra letter. We use six copies of each such ``plain''
letter. These copies are obtained by priming and double-priming letters,
and by overlining. Hence the six copies of $\tta$ are
$\tta,\tta',\tta'',\overline{\tta},\overline{\tta'},\overline{\tta''}$. As expected, for a ``plain'' word
(or alphabet) $x$, we write $x'$ and $\overline{x}$ to denote a version of $x$
obtained by priming (respectively, overlining) all its letters. Formally,
letting $\Upsilon_\xs$ being short for $\Upsilon\cup\{\xs\}$, one has
$\Sigma=\Gamma\egdef \Upsilon_\xs\cup \Upsilon'_\xs\cup \Upsilon''_\xs\cup
\overline{\Upsilon_\xs}\cup \overline{\Upsilon'_\xs}\cup \overline{\Upsilon''_\xs} $.

We define and explain the reduction by running it on the following example:
\begin{gather}
\label{eq-S-exmp}
\tag{$S_{\text{exmp}}$}
\Upsilon=\{\tta,\ttb,\ttc\}
\text{ and }
\Delta=\{\tta\ttb\rightarrow \ttb\ttc,\; \ttc\ttc\rightarrow \tta\tta\}.
\end{gather}
Assume that $\tta\ttb\ttc\in P_1$ and $\ttb\tta\tta\in P_2$. Then $P_1\rews P_2$ since
$\tta\ttb\ttc\rews \ttb\tta\tta$ as witnessed by the following (even-length)
derivation $\pi$ = ``$\tta\ttb\ttc \rew \ttb\ttc\ttc \rew \ttb\tta\tta$''.
In our reduction, a rewrite step like ``$\tta\ttb\ttc \rew \ttb\ttc\ttc$'' appears in the PEP
solution $\sigma$ as the letter-by-letter interleaving $\tta\overline{\ttb}\ttb\overline{\ttc}\ttc\overline{\ttc}$, denoted
$\tta\ttb\ttc\inl \ttb\ttc\ttc$, of a plain string and an overlined copy of a same-length string.

Write $T_\btr(\Delta)$, or just $T_\btr$ for short, for the set of all
$x\inl y$ such that $x\rew y$. Obviously, and since we are dealing
with length-preserving systems, $T_\btr$ is a regular language, as seen by
writing it as
$
T_\btr =  \bigl( \sum_{a\in\Upsilon}a\overline{a} \bigr)^*
 .	 \bigl\{ l\inl r ~|~ l\rightarrow r\in\Delta \bigr\} .
	 \bigl( \sum_{a\in\Upsilon}a\overline{a} \bigr)^*
$,
where $\{l\inl r ~|~ l\rightarrow r\in\Delta\}$ is a finite, hence regular, language.

$T_\btr$ accounts for odd-numbered steps. Symmetrically, for even-numbered steps
like $\ttb\ttc\ttc\rew \ttb\tta\tta$ in $\pi$ above, we use $\ttb\overline{\ttb}\tta\overline{\ttc}\tta\overline{\ttc}$, i.e., $\ttb\tta\tta\inl \ttb\ttc\ttc$.
Here too $T_\btl\egdef\{y\inl x ~|~x\rew y\}$ is regular.
Finally, a derivation $\pi$ of the general form
\[x_0\rew x_1\rew x_2 \ldots
\rew x_{2k},\]
where $K\egdef \size{x_0}=\ldots=\size{x_{2k}}$, is encoded
as a solution $\sigma_\pi$ of the form
$
\sigma_\pi =
\rho_0 \sigma_1\rho_1 \sigma_2 \ldots \rho_{2k-1} \sigma_{2k}\rho_{2k}
$ that alternates between the encodings of steps (the $\sigma_i$'s) in
$T_\btr\cup T_\btl$, and \emph{fillers}, (the $\rho_i$'s) defined as follows:
\begin{xalignat*}{1}
%\label{eq-def-sigmai&rhoi}
\sigma_i &\egdef
\left\{\begin{array}{ll}
x_{i-1}\inl x_i &\text{for odd $i$}\:,\\[.5em]
x_i\inl x_{i-1} &\text{for even $i$}\:,
\end{array}\right. %\}
\end{xalignat*}
\begin{xalignat*}{2}
%\label{eq-def-sigmai&rhoi}
\begin{array}{l}
\rho_0\egdef x''_0\inl \xs''{}^K\:,\\[.5em]
\rho_{2k}\egdef x''_{2k}\inl \xs''{}^K\:,
\end{array}&
&
\rho_i &\egdef
\left\{\begin{array}{ll}
\xs'{}^K\inl x'_i&\text{for odd $i$}\:,\\[.5em]
x'_i\inl \xs'{}^K &\text{for even $i\not=0,2k$}\:.
\end{array}\right. %\}
\end{xalignat*}
Note that the extremal fillers $\rho_0$ and $\rho_{2k}$ use double-primed
letters, when the internal fillers use primed letters.
Continuing our example, the $\sigma_\pi$ associated with the
derivation $\tta\ttb\ttc \rew \ttb\ttc\ttc \rew \ttb\tta\tta$ is
\[
\sigma_\pi =
\ubracew{\tta''\overline{\xs''}\ttb''\overline{\xs''}\ttc''\overline{\xs''}}{\tta''\ttb''\ttc''\inl \xs''\xs''\xs''}
\;\;
\ubracew{\tta\overline{\ttb}\ttb\overline{\ttc}\ttc\overline{\ttc}}{\tta\ttb\ttc\inl \ttb\ttc\ttc}
\;\;
\ubracew{\xs'\overline{\ttb'}\xs'\overline{\ttc'}\xs'\overline{\ttc'}}{\xs'\xs'\xs'\inl \ttb'\ttc'\ttc'}
\;\;
\ubracew{\ttb\overline{\ttb}\tta\overline{\ttc}\tta\overline{\ttc}}{\ttb\tta\tta\inl \ttb\ttc\ttc}
\;\;
\ubracew{\ttb''\overline{\xs''}\tta''\overline{\xs''}\tta''\overline{\xs''}}{\ttb''\tta''\tta''\inl \xs''\xs''\xs''}
\:.
\]
The point with primed and double-primed copies is that $u$ and $v$
associate them with different images. Precisely, we define
\begin{xalignat*}{5}
u(a) &=a ,
&
u(a') &=\xs ,
&
u(\xs') &=\xs ,
&
u(a'') &=\epsilon ,
&
u(\xs'') &=\epsilon ,
\\
v(a)&=\xs ,
&
v(a')&=a ,
&
v(\xs')&=w_\Upsilon ,
&
v(a'')&=a ,
&
v(\xs'')&=w_\Upsilon ,
\end{xalignat*}
where $a$ is any letter in $\Upsilon$, and where $w_\Upsilon$ is a word
listing all letters in $\Upsilon$. E.g., $w_{\{\tta,\ttb,\ttc\}}=\tta\ttb\ttc$ in our
running example.
The  extremal fillers use special double-primed letters  because we want
$u(\rho_0) = u(\rho_{2k}) = \epsilon$ (while $v$ behaves the same on primed
and double-primed letters).
Finally, overlining is preserved by $u$ and $v$:
$u(\overline{x}) \egdef \overline{u(x)}$ and $v(\overline{x}) \egdef
\overline{v(x)}$.
\iffalse
\begin{xalignat*}{2}
u(\overline{x}) &= \overline{u(x)} ,
&
v(\overline{x}) &= \overline{v(x)} \:.
\end{xalignat*}
\fi
%

This ensures that, for $i>0$, $u(\sigma_i)\subword v(\rho_{i-1})$ and
$u(\rho_i)\subword v(\sigma_i)$, so that a $\sigma_\pi$ constructed as
above is a direct solution. It also ensures $u(\sigma_i)\subword v(\rho_i)$
and $u(\rho_{i-1})\subword v(\sigma_i)$ for all $i>0$, so that $\sigma_\pi$
is also a codirect solution. One can check it on our running example by
writing $u(\sigma_\pi)$ and $v(\sigma_\pi)$ alongside:
\[
\begin{array}{rccccc}
\sigma_\pi\; \!=\! &
\;\obracew{\tta''\overline{\xs''}\ttb''\overline{\xs''}\ttc''\overline{\xs''}}{\rho_0}\;
&
\;\obracew{\tta\overline{\ttb}\ttb\overline{\ttc}\ttc\overline{\ttc}}{\sigma_1}\;
&
\;\obracew{\xs'\overline{\ttb'}\xs'\overline{\ttc'}\xs'\overline{\ttc'}}{\rho_1}\;
&
\;\obracew{\ttb\overline{\ttb}\tta\overline{\ttc}\tta\overline{\ttc}}{\sigma_2}\;
&
\;\obracew{\ttb''\overline{\xs''}\tta''\overline{\xs''}\tta''\overline{\xs''}}{\rho_2}\;
\\[.4em]
\hline
\\[-.4em]
\!u(\sigma_\pi)\; \!=\! &
&
\;{\tta\overline{\ttb}\ttb\overline{\ttc}\ttc\overline{\ttc}}\;
&
\;{\xs\overline{\xs}\xs\overline{\xs}\xs\overline{\xs}}\;
&
\;{\ttb\overline{\ttb}\tta\overline{\ttc}\tta\overline{\ttc}}\;
&
\\[.4em]
\!v(\sigma_\pi)\; \!=\! &
\tta\,\overline{\tta\ttb\ttc}\,\ttb\,\overline{\tta\ttb\ttc}\,\ttc\,\overline{\tta\ttb\ttc}
&
\;{\xs\overline{\xs}\xs\overline{\xs}\xs\overline{\xs}}\;
&
\tta\ttb\ttc\,\overline{\ttb}\,\tta\ttb\ttc\,\overline{\ttc}\,\tta\ttb\ttc\,\overline{\ttc}
&
\;{\xs\overline{\xs}\xs\overline{\xs}\xs\overline{\xs}}\;
&
\ttb\,\overline{\tta\ttb\ttc}\,\tta\,\overline{\tta\ttb\ttc}\,\tta\,\overline{\tta\ttb\ttc}
\end{array}
\]

There remains to define $R$. Since $\rho_0\in
\bigl(\Upsilon''\overline{\xs''}\bigr)^+$, since $\sigma_i\in T_\btr$ for
odd $i$, etc., we let
\begin{gather*}
%\label{eq-def-R}
R \:\egdef\:
\bigl(\Upsilon''\overline{\xs''}\bigr)^+
. T_\btr^{\cap P_1} .
\bigl(\xs'\overline{\Upsilon'}\bigr)^+
.
\Bigl(
T_\btl
. \bigl(\Upsilon'\overline{\xs'}\bigr)^+
. T_\btr .
\bigl(\xs'\overline{\Upsilon'}\bigr)^+
\Bigr)^*
. T_\btl^{\cap P_2} .
\bigl(\Upsilon''\overline{\xs''}\bigr)^+
\:,
\end{gather*}
where $T_\btr^{\cap P_1}\egdef \{x\inl y~|~ x\rew y \wedge x\in
P_1\}=T_\btr\cap \{x\inl y~|~x\in P_1\wedge\size{x}=\size{y}\}$ is
clearly regular when $P_1$ is, and similarly for $T_\btl^{\cap P_2}\egdef
\{y\inl x~|~ x\rew y \wedge y\in P_2\}$. Since $\sigma_\pi\in
R$ when $\pi$ is an even-length derivation from $P_1$ to $P_2$, we deduce
that the left-to-right implication in \eqref{eq-reduction-spec} holds.
\\

We now prove the right-to-left implication,
which concludes the proof of Theorem~\ref{theo-undec-pepdcd}.

Assume that there is a $\sigma\in R$ such that $u(\tau)\subword v(\tau)$ and
$u(\tau')\subword v(\tau')$ for all decompositions $\sigma=\tau\tau'$. By
definition of $R$, $\sigma$
must be of the form
\[
\sigma = \rho_0\sigma_1\rho_1(\sigma_2\rho_2\sigma_3\rho_3)\ldots
(\ldots\sigma_{2k-1}\rho_{2k-1})\sigma_{2k}\rho_{2k}
\]
for some $k>0$, with
$\rho_0\in\bigl(\Upsilon'' \overline{\xs''}\bigr)^+$, with $\sigma_i\in T_\btr$ for odd
$i$ and $\sigma_i\in T_\btl$ for even $i$, etc. These $4k+1$ non-empty factors,
$(\sigma_i)_{1\leq i\leq 2k}$ and $(\rho_i)_{0\leq i\leq 2k}$, are called the
``\emph{segments}'' of $\sigma$, and numbered $s_0,\ldots,s_{4k}$ in order.

\begin{lem}
\label{lem-aux1}
$u(s_p)\subword v(s_{p-1})$ and $u(s_{p-1})\subword v(s_p)$
for all $p=1,\ldots,4k$.
\end{lem}
\begin{proof}
First note that the definition of $u$ and $v$ ensures that $u(s_p)$ and
$v(s_p)$ use disjoint alphabets. More precisely, all $u(\sigma_i)$'s and
$v(\rho_i)$'s are in $(\Upsilon\overline{\Upsilon})^*$, while the
$v(\sigma_i)$'s and the $u(\rho_i)$'s are in $(\xs\overline{\xs})^*$, with the
special case that $u(\rho_0)=u(\rho_{2k})=\epsilon$ since $\rho_0$ and
$\rho_{2k}$ are made of double-primed letters.

Since $\sigma$ is a direct solution, $u(s_0\ldots
s_p)\subword v(s_0\ldots s_p)$ for any $p$, and even
\begin{gather}
\tag{$A_p$}
\label{eq-leftmost}
	      u(s_0\ldots s_p)\subword v(s_0\ldots s_{p-1}),
\end{gather}
since $v(s_p)$ has no letter in common with $u(s_p)$. We now claim that,
for all $p=1,\ldots,4k$
\begin{gather}
\tag{$B_p$}
\label{eq-leftmost-not}
	 u(s_0s_1\ldots s_p)\not\subword v(s_0s_1\ldots s_{p-2}),
\end{gather}
as we prove by induction on $p$. For the base case, $p=1$, the claim is
just the obvious $u(s_0s_1)\not\subword \epsilon$. For the inductive case
$p>1$, one combines $u(s_0\ldots s_{p-1})\not\subword v(s_0\ldots s_{p-3})$
(ind.\	hyp.) with $u(s_p)\not\subword v(s_{p-2})$ (different alphabets) and
gets $u(s_0\ldots s_p)\not\subword v(s_0\ldots s_{p-2})$.

We now
combine $(A_p)$, i.e., $u(s_0\ldots s_p)\subword
v(s_0\ldots s_{p-1})$, and $(B_{p-1})$,  i.e., $u(s_0s_1\ldots s_{p-1})\not\subword v(s_0s_1\ldots
s_{p-3})$,  yielding $u(s_p)\subword v(s_{p-2}s_{p-1})$, hence $u(s_p)\subword
v(s_{p-1})$ since $u(s_p)$ and $v(s_{p-2})$ share no letter: we have proved one
half of the Lemma.
The other half is proved symmetrically, using the fact that
$\sigma$ is also a codirect solution.
\qed
\end{proof}
\begin{lem}
\label{lem-aux2}
$\size{s_1}=\size{s_2}=\ldots=\size{s_{4k-1}}$.
\end{lem}
\begin{proof}
For any $p$ with $0<p<4k$, $u(s_p)\subword v(s_{p-1})$
(Lemma~\ref{lem-aux1}) implies $\size{s_p}\leq\size{s_{p-1}}$, as can
easily be seen either when $s_p$ is some $x\inl y$ or when $s_p$ is some
filler like $\xs'{}^L\inl x'$. Thus $\size{s_0}\geq
\size{s_1}\geq\cdots\geq \size{s_{4k-1}}$. Similarly, the other half of
Lemma~\ref{lem-aux1}, i.e., $u(s_{p-1})\subword v(s_p)$, entails
$\size{s_1}\leq \size{s_2}\leq \cdots\leq \size{s_{4k}}$.
\qed
\end{proof}

Now pick any $i\in\{1,\ldots,2k\}$. If $i$ is odd, then by definition of $R$, $\sigma_i\in
T_\btr$ is some $x_{i-1}\inl y_i$ with $x_{i-1}\rew y_i$ and $\sigma_{i+1}\in
T_\btl$ is some $y_{i+1}\inl x_i$ with $x_i\rew y_{i+1}$.
Furthermore, $\rho_i$ is some $\xs'{}^{\size{z_i}}\inl z'_i$. With
Lemma~\ref{lem-aux1}, we deduce $y_i\subword z_i$ and $x_i\subword
z_i$. With Lemma~\ref{lem-aux2}, we further deduce
$\size{y_i}=\size{z_i}=\size{x_i}$, hence $y_i=x_i$. A similar
reasoning shows that $y_i=x_i$ also holds when $i$ is even, so that the
steps $x_{i-1}\rew y_i$ can be chained. Finally,
we deduce from $\sigma$ the existence of a derivation $x_0\rew x_1\rew
\cdots \rew x_{2k}$. Since $\sigma_0\in T_\btr^{\cap P_1}$ and $\sigma_{2k}\in
T_\btl^{\cap P_2}$, we further deduce $x_0\in P_1$ and $x_{2k}\in P_2$.
Hence the existence of $\sigma$ entails $P_1\rewn{2k} P_2$, which concludes the
proof.

%%% Local Variables:
%%% fill-column: 75
%%% ispell-check-comments: nil
%%% Local IspellDict: "english"
%%% End:

% LocalWords:  DCFL xy codirectness codirect rl overlining exmp overlined eq CP
% LocalWords:  rccccc codirected fsttcs

%%% File: sec-concl.tex -*-LaTeX-*-

\section{Concluding remarks}
%===========================
\label{sec-concl}

We introduced partial directness in Post Embedding Problems and proved the
decidability of $\PEPpcod$ and $\PEPpd$ by showing that an instance has a solution if, and
only if, it has a solution of length bounded by a computable function of
the input. 
(Furthermore, from Theorem~\ref{theo-short}, one may directly derive upper bounds on
the complexity of $\PEPpcod$ and $\PEPpd$ using the bounds on the Length Function
$H$ provided in~\cite{SS-icalp11}.)

This generalizes and simplifies earlier proofs for $\PEP$ and
$\PEPd$. The added generality is non-trivial and leads to decidability for
UCST, or UCS (that is, unidirectional channel systems) extended with
tests~\cite{JKS-tcs2012}. The simplification lets us deal smoothly with
counting or universal versions of the problem. Finally, we showed that
\emph{combining} directness and codirectness constraints leads to
undecidability.

%%% Local Variables:
%%% ispell-check-comments: nil
%%% Local IspellDict: "english"
%%% fill-column: 75
%%% End:

% LocalWords:  codirectness UCS LCS UCST

%% \begin{acknowledgements}
%%  \ldots
%% \end{acknowledgements}

%\nocite{*} % == test biblio
%\bibliographystyle{spmpsci}
%\bibliographystyle{spbasic}
{\small
\bibliographystyle{plain}
\bibliography{lcs}
}
%\appendix
%\input{appendix}

%%% Local Variables:
%%% fill-column: 75
%%% ispell-check-comments: nil
%%% Local IspellDict: "english"
%%% End:

\end{document}